# HDH: A Python Library for Hypergraph-Based Distributed Quantum Computing Partitioning

**Maria Gragera Garces** [1]

**1** University of Edinburgh

## Summary

Today's quantum computers are limited by how many qubits a single device can hold. Distributed quantum computing (DQC) works around this by linking multiple smaller devices together so they can jointly run computations too large for any one of them alone, which first requires deciding how to split, or partition, a computation across those devices.

`HDH` (Hybrid Dependency Hypergraphs) is a Python library that gives researchers a common representation to develop and compare partitioning strategies against. It converts a quantum computation — expressed as a circuit, a measurement-based quantum computing (MBQC) pattern, a quantum walk, or a quantum cellular automaton — into a hypergraph that captures every way the computation could be split across devices, including hard constraints such as per-device qubit limits that prior abstractions treat as soft penalties rather than requirements. Researchers can run their own partitioning heuristics directly on this representation, or use `HDH`'s built-in capacity-aware baseline, and compare results on a consistent, model-agnostic footing. `HDH` also imports circuits from popular quantum SDKs (Qiskit, Cirq, PennyLane, Amazon Braket) and exports back to Qiskit and PennyLane, so partitioned results can be turned back into runnable circuits.

## Statement of need

HDHs were originally proposed by Gragera Garces et al. (2025) as a unifying approach to quantum distribution, extending the hypergraph abstraction that Andrés-Martínez & Heunen (2019) introduced for partitioning across devices. Various partitioning strategies have since been proposed building on that earlier abstraction (Clark et al., 2023; Escofet et al., 2023; Sundaram & Gupta, 2023), but many are tested on inconsistent hypergraph abstractions, hindering cross-partitioner comparison and improvement.

An open-source, model-agnostic abstraction enables fair cross-comparison of partitioning strategies, and extends it beyond the circuit model, a current blind spot in DQC research. `HDH` is designed for distributed quantum architecture researchers and for compiler developers who need a model-agnostic distribution layer.

## State of the field

Existing DQC approaches abstract computations to hypergraphs which are then partitioned, using balanced hypergraph partitioning solvers such as KaHyPar (Schlag et al., 2023).

This framing has two fundamental limitations:

1. It reduces distribution to a balanced partitioning problem that ignores a hard physical constraint: individual QPUs have fixed qubit capacities, and a valid distribution must respect these limits strictly rather than treating them as soft penalties.

2. Existing hypergraph abstractions are model-specific and encode only a subset of possible partition cuts, meaning partitioning strategies are routinely evaluated on inconsistent abstractions, making cross-comparison unreliable and hindering the systematic development of improved heuristics.

Libraries for distributed quantum computing exist, such as DISQCO (Burt et al., 2026), Qdislib (Tejedor et al., 2025), Optyx (Kupper et al., 2025) and DC-MBQC (Xue et al., 2026), but they implement end-to-end distribution pipelines rather than exposing the underlying abstraction as a research tool. No existing library provides a model-agnostic hypergraph abstraction designed to enable the development and fair comparison of partitioning heuristics, which is the role `HDH` is built to fill.

Quantum compilation frameworks like Qiskit (Javadi-Abhari et al., 2024), Cirq (Cirq Developers, 2026), and PennyLane (Bergholm et al., 2018) provide circuit optimization and device mapping, but not model-agnostic abstractions for distributed quantum computing. `HDH` is compatible with these SDKs and adds to them rather than replacing them.

# Software design

The central design decision in `HDH` was to separate the abstraction layer from the partitioning layer. Rather than building a monolithic tool that both constructs hypergraphs and partitions them, `HDH` exposes the HDH as a first-class data structure that any downstream partitioner can consume. This makes the library useful both as a standalone research tool and as a substrate for third-party heuristics.

`HDH` uses hypergraphs because quantum operations are not limited to two qubits: a Toffoli gate, for instance, acts on three at once. An ordinary graph is the special case in which every edge joins exactly two nodes, so a multi-qubit operation must be split into several pairwise edges. A partition that cuts through it can then be charged once per severed edge, although the operation is a single unit that is cut, and communicated, as a whole. A hyperedge keeps it as one object, so the cut cost matches the communication cost.

Two further design choices trade added complexity for partitioning flexibility. First, HDHs represent each qubit's state at every timestep as a separate node rather than a single node per logical qubit, and partitioning operates over these timestepped nodes rather than whole qubits. This allows a single qubit's history to be split across devices at whichever point in the computation makes the split cheapest, with its state teleported between them at that boundary, rather than committing the qubit to one device for the circuit's full duration. Second, a multi-qubit gate is represented not as one hyperedge but three: one linking each qubit's pre-gate state to an intermediate state, one spanning all involved qubits at that intermediate point, and one linking each qubit's post-gate state onward. This staging is what allows a partitioner to cut through a multi-qubit gate at either boundary — modelling either a non-local gate executed over the network or a teleportation of one qubit's state immediately before or after the gate — instead of being limited to the coarser choice of which side of a single gate-edge to place a device boundary on.

The library includes a capacity-aware greedy heuristic as a built-in baseline. Existing DQC research typically benchmarks against KaHyPar (Schlag et al., 2023), a general-purpose partitioner not designed for quantum hardware constraints. The included heuristic is simpler, but treats per-device qubit capacity as a hard constraint and returns a feasible assignment whenever one exists, making it a more appropriate DQC baseline and a starting point for improved strategies.

## From computational models to HDHs

HDHs use the following notation to describe quantum workload dependencies, including predicted elements that represent potential future state transformations based on classical measurement outcomes:

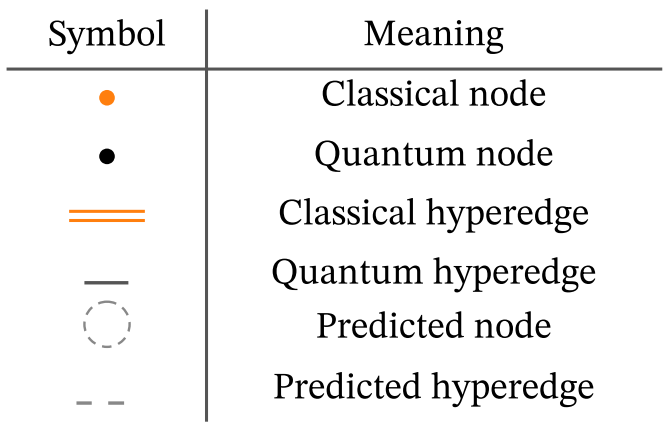

**Figure 1:** HDH symbol legend.

Mapping a quantum workload such as a circuit to an HDH applies correspondences between model elements and hypergraph motifs. Model-specific classes such as `Circuit` perform these conversions; any class with a `build_hdh()` method satisfies the library's `Model` protocol and is checked by the same conformance tests:

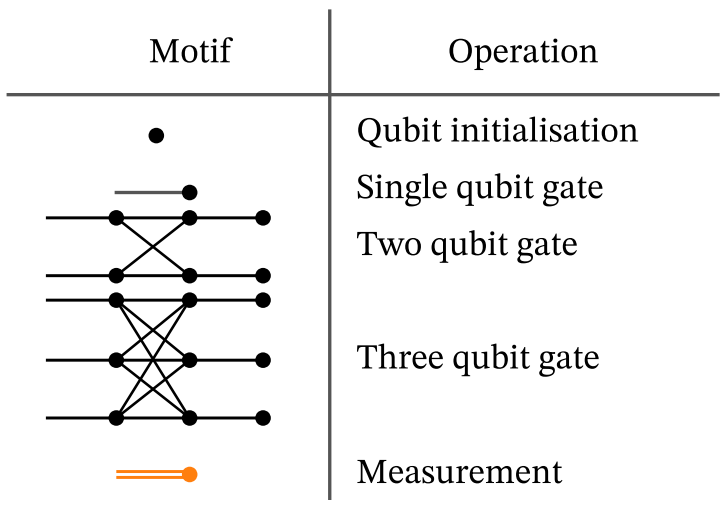

**Figure 2:** Circuit to HDH mapping table.

In the context of DQC, entangling operations in a model can be made non-local (namely non-local gates) and thus partitioned through a quantum network via quantum communication primitives (Wu et al., 2022). Alternatively, qubit states can be individually forwarded through teleportation protocols (Zomorodi-Moghadam et al., 2018). Because HDHs represent both cut types within a single structure, a partitioner is free to combine them within one workload — e.g. keeping a qubit local to a device via non-local gates during one phase of a computation, then teleporting it elsewhere for a later phase — rather than being restricted to whichever single strategy the chosen abstraction happens to support.

Unlike prior abstractions, which represent only non-local gates or only teleportation, HDHs capture both (Figure 3).

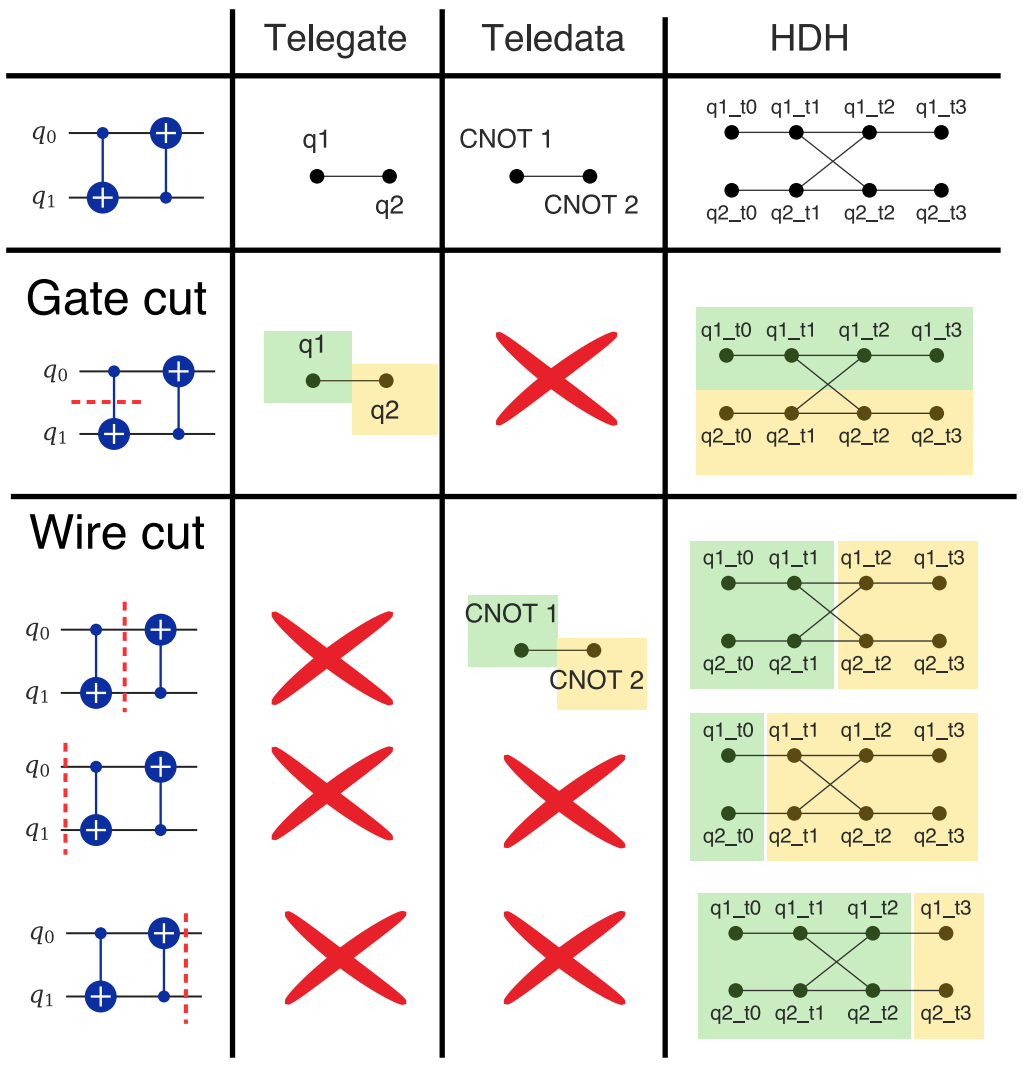

**Figure 3:** Table showing HDH expressivity.

Figure 4 shows the HDH of a six-qubit circuit with a Toffoli gate, a classically conditioned gate and mid-circuit measurements, drawn as a graph. Each gate contributes the hyperedge of its state transformation plus the preceding and following hyperedges that allow pre- and post-teleportation, nodes are possible state transformations rather than qubits or operations, and classical data flow (orange) is part of the hypergraph.

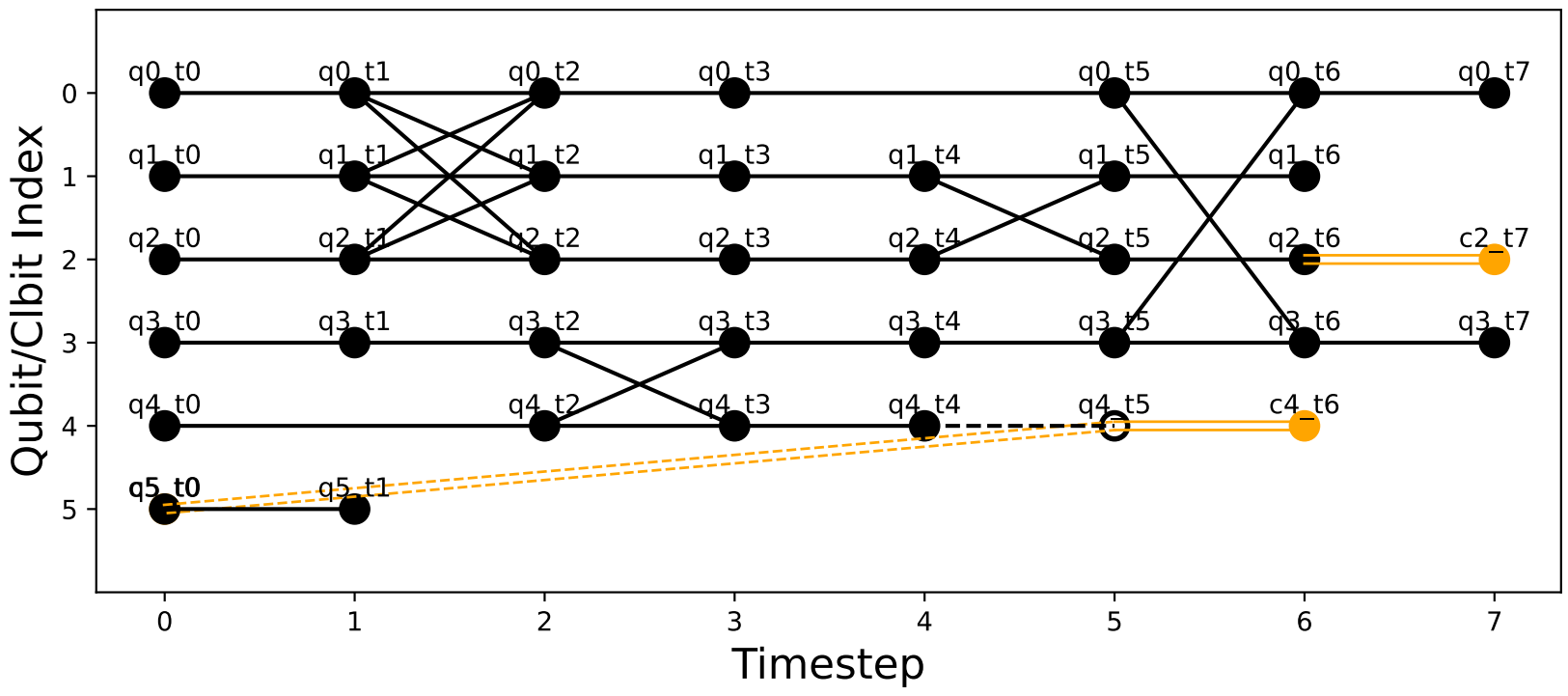

**Figure 4:** Example circuit and its HDH representation.

## Research impact statement

HDH's capacity-aware partitioner is proven optimal on 20 of 24 instances (83%) and averages $1.03\times$ the optimal cut cost (worst case $1.33\times$), measured against exhaustive branch-and-bound search over MQT Bench circuits (Quetschlich et al., 2023) at the tightest feasible capacity (three devices, network overhead 1). It also returns a feasible assignment whenever one exists, which the balance-based partitioners used as DQC baselines do not guarantee.

Carrying teledata and telegate cuts in one structure means the combined formulation automatically matches the best single-mode strategy, whichever that is for a given interaction pattern. Where a qubit interacts with one group of qubits per phase, shifting groups between phases, exhaustive search over two, three and four phases (one, two and three shifts) gives cut costs of 1, 2 and 3, one per shift, against 3, 3 and 6 for the qubit-level formulation used by prior hypergraph approaches ($1.5$–$3\times$ less), matching the teleportation-only formulation throughout. On 30 feasible MQT Bench instances (12 circuits, three device settings), where no such shift occurs, a greedy placer on the combined formulation ties the qubit-level one on every instance, while assigning $8.4\times$ as many placement units per circuit (mean over 12 circuits; $3.6$–$14.7\times$). A simple greedy does not always reach the combined optimum, though: on 3 of those 30 instances, a greedy restricted to teleportation cuts did better.

What has no counterpart elsewhere is that the same partitioner runs unmodified over circuits, MBQC patterns, quantum walks, and quantum cellular automata, exercised across all four by the test suite. Comparing distribution overhead between computational models, which a circuit-only library cannot do, is here a matter of swapping the workload builder; a first such study is being developed by the authors.

Every quantitative claim in this section regenerates from `benchmarks/`.

Early community engagement has been encouraging. The project was presented as a poster at SIGCOMM 2025 (Gragera Garces et al., 2025), has received funding through the Unitary Fund microgrant program and has already seen external contributors (acknowledged below). We are also in discussion with companies in the distributed quantum computing space about integrating the library into their stack.

# AI usage disclosure

Claude was used during both library development and paper writing. Claude Opus 5 and Claude Sonnet 5.5 (through Claude Code) were the models used, for both the library and the paper.

In the library, Claude generated initial draft code implementations that were subsequently rewritten by the author. It also assisted in producing unit tests, which were validated against expected behaviour across both passing and failing scenarios. These were not always fully re-written but they were revised and thoroughly tested. Additionally, AI was used to generate inline code comments throughout the library, with the aim of improving readability for contributors and users who may check the source code.

In the paper, Claude was used to assist with wording and polish.

All AI-generated content (both code and text) was reviewed or modified (as per the above descriptions) by the author before inclusion in the present version.

# Acknowledgements

We acknowledge contributions from Joseph Tedds, Manuel Alejandro, and Alessandro Cosentino.

We thank Unitary Fund for supporting this project through their quantum microgrant program.

The work of the author is supported by the EPSRC UK Quantum Technologies Programme under grant EP/T001062/1 and VeriQloud.